\documentstyle[aps]{revtex}
\author{
Antonio Coniglio and Mario Nicodemi 
}
\address{
\vspace{0.2cm}
Dipartimento di Fisica, Universit\'a di Napoli ``Federico II'',
INFM and INFN Sezione di Napoli \\
Mostra d'Oltremare, Pad. 19, 80125 Napoli, Italy\\
}

\title{ Scaling properties in off equilibrium dynamical processes}

\date{\today}

\begin{document}

\maketitle
\bigskip
\begin{abstract}
In the present paper, we analyze the consequences of scaling hypotheses on 
dynamic functions, as two times correlations $C(t,t')$.
We show, under general conditions, that $C(t,t')$ must obey 
the following scaling behavior 
$C(t,t') = \phi_1(t)^{f(\beta)}{\cal{S}}(\beta)$, 
where the scaling variable is $\beta=\beta(\phi_1(t')/\phi_1(t))$ 
and $\phi_1(t')$, $\phi_1(t)$ two undetermined functions.
The presence of a non constant exponent $f(\beta)$ signals the 
appearance of multiscaling properties in the dynamics. 
\end{abstract}

\bigskip

\section{Introduction}
The introduction of scaling concepts to describe equilibrium and 
off equilibrium dynamics in statistical mechanics was originally 
motivated by experimental and simulation data about, for instance, 
structure factor, pair correlation functions, response functions. 
Actually, the study of several classes of materials with complex dynamical 
properties as magnets, polymers, glasses, and several other thermal systems, 
and even non-thermal systems as granular media, has shown 
the presence of some general scaling features \cite{HH,Bray,BCKM}. 
In order to formulate a coherent scaling approach to the dynamics 
of systems out of
equilibrium, in this paper we resort to a general scheme developed in 1971
\cite{CM}. This approach also reproduces as a particular case the 
multifractal and the multiscaling formalism, 
which has been applied to a large variety of 
phenomena such as turbulence, random resistor networks, self organized 
criticality, spinodal decomposition and many more
\cite{mandel,turbulence,amitrano,Kadanoff_ms,deArcangelis,halsey,PV,Stanley,CZ}.
The general scaling formulation applied to systems out of equilibrium
stems from the hypothesis of 
invariance of two time functions, as autocorrelation functions, 
under a general scaling transformation with the only requirement
that the transformation obeys group properties. 

For definiteness let's consider a two time correlation function, 
$C(t,t')$, which for example could be the density-density 
autocorrelation function in a supercooled liquid or the spin-spin 
correlation function in a magnetic system. 
We suppose that the system is prepared at time $t=0$ and is probed at 
two subsequent time $t'$ and $t$. When the system is 
out of equilibrium, $C(t,t')$ generally depends on both $t'$ and $t$.
Whenever the relaxation characteristic time, $\tau$, is very large or 
infinite, we may make the following asymptotic scaling ansatz 
valid for $t$ and $t'$ large but smaller than $\tau$: 
by rescaling the system lengths by a factor $l$, 
if $t$ and $t'$ are opportunely rescaled, we may expect that 
the autocorrelation function scales as $l^{f(t,t')}$, where the exponent 
$f(t,t')$ is in general dependent on $t$ and $t'$ \cite{nota1}. 
To be precise we assume that the function $C(t,t')$ has the following 
general scaling property
\begin{equation}
C(\tilde{t},\tilde{t'})=l^{-f(t,t')}C(t,t') 
\label{gen_sca_C}
\end{equation}
under a general time rescaling {\em ``non mixing"} 
transformation, which satisfies group rules, as 
\begin{equation}
\tilde{t}=F_1(t,l) ~~~~~~  \tilde{t'}=F_2(t',l)
\label{gen_tra_t}
\end{equation}
with the condition that $F_i(x,1)=x$ ($i=1,2$).
The above transformations are ``non mixing" in the sense that 
$F_1$ (resp. $F_2$) depends only on $t$ (resp. $t'$).
The requirement that the transformation obeys group properties imposes 
some constraints on the functions $C(t,t')$ and $f(t,t')$.
Interestingly, under these assumptions, we find that $C(t,t')$ 
can be synthetically expressed in the following way:
\begin{equation}
C(t,t') = \phi_1(t)^{f(\beta)}{\cal{S}}(\beta),
\label{sca_fin}
\end{equation}
where the scaling variable, $\beta$, has the following form:
\begin{equation}
\beta=\phi_2(t')/\phi_1(t) ~~.
\label{power_sca}
\end{equation}
Here the $\phi_i$ ($i=1,2$) are two unknown functions fixed by the 
transformations given in eq.~(\ref{gen_tra_t}). 
Eq.~(\ref{sca_fin}) in the particular case $f(\beta)=0$ was obtained in 
Ref.\cite{Cu_Ku} using different arguments. 
Notice that, whenever $f$ is not a constant, a ``multiscaling" dynamical 
behavior is found in the dynamics, an interesting issue to check  
in models as well as experiments and simulations.

\section{The general ``non mixing" case}

In what follows we give a demonstration of what summarized above. 
As shown in Ref.\cite{CM}, the general transformations 
of eq.~(\ref{gen_tra_t}) implies that exist a couple of functions, 
$\phi_1(t)$ and $\phi_2(t')$, which under rescaling
exhibits the following property:
\begin{equation}
\phi_1(\tilde{t})=\phi_1(t)/l ~~~~~~  \phi_2(\tilde{t'})=\phi_2(t')/l
\label{sca_fun}
\end{equation}

These equations state that the ``true" scaling variables are the $\phi_i$'s, 
and that, whenever the functions $\phi_i$'s are {\em invertible}, 
eq.s~(\ref{gen_tra_t}) can be expressed in the following way:
\begin{equation}
\tilde{t}\equiv F_1(t,l)=\phi_1^{-1}(\phi_1(t)/l) ~~~~~~  
\tilde{t'}\equiv F_2(t',l)=\phi_2^{-1}(\phi_2(t')/l)
\label{gen_tra_phi}
\end{equation}
where $\phi_i^{-1}$ is the inverse function of $\phi_i$ 
(i.e., $\phi_i^{-1}(\phi_i(x))=x$).

Let's now study how the above group properties influence the 
structure of the function $C(t,t')$. 
The group rules impose that if we scale $t$ and $t'$ by a factor $l_1$ and 
later by a factor $l_2$ this should be equivalent to rescale them by a factor 
$l_1\cdot l_2$. More formally, we can express this condition as:
\begin{equation}
C(F_1(F_1(t,l_1),l_2),F_2(F_2(t',l_1),l_2))= 
(l_1l_2)^{-f(t,t')} C(t,t') ~~ . 
\label{group_on_C}
\end{equation}
Substituting eq.s~(\ref{gen_sca_C}) and (\ref{gen_tra_t}) in the above 
relation, one is led to a simple equation which states that:
\begin{equation}
\frac{d}{d l} f(t(l),t'(l)) = 0
\label{eq_f}
\end{equation}
where, by definition, $t(l)=F_1(t,l)$ and $t'(l)=F_2(t,l)$. 
By inserting eq.(\ref{gen_tra_phi}) in eq.~(\ref{eq_f}), one finds that 
$f(\phi_1^{-1}(\phi_1/l),\phi_2^{-1}(\phi_2/l))=
f(\phi_1^{-1}(\phi_1/1),\phi_2^{-1}(\phi_2/1))$, i.e., we have that 
$f(\phi_1^{-1}(\phi_1/l),\phi_2^{-1}(\phi_2/l))
=f(t,t')$. Now, by taking $l=\phi_1$ we obtain that 
$f(t,t')=f(\phi_1^{-1}(1),\phi_2^{-1}(\phi_2/\phi_1))$, that is to say
$f(t,t')=f(\phi_2(t')/\phi_1(t))$.

Analogously, by inserting eq.(\ref{gen_tra_phi}) in eq.~(\ref{gen_sca_C}), 
and choosing $l=\phi_1$, we find the scaling form for $C(t,t')$ 
that we anticipated in eq.~(\ref{sca_fin}) above. 
In such a way we also individuate the scaling function ${\cal{S}}$: 
${\cal{S}}(x)=C(\phi_1^{-1}(1),\phi_2^{-1}(x))$. 

Thus we proved that in presence of scaling properties 
as those written in eq.s~(\ref{gen_sca_C}) and (\ref{gen_tra_t}), 
the asymptotic functional form of the scaling of $C(t,t')$ is 
characterized by the asymptotic behavior of the ``true" scaling 
variables $\phi_1$ and $\phi_2$, as written in the 
general result of eq.~(\ref{sca_fin}). 

For sake of clarity we have dealt with a two variables function, 
$C(t,t')$, but analogous properties may be proven for a 
many variables function, $C(t_1,t_2,...,t_n)$. In this case, if the 
generic variable undergoes a scale transformation 
$\tilde{t_i}=F_i(t_i,l)$ ($i\in\{1,...,n\}$), we have:
\begin{equation}
C(t_1,t_2,...,t_n) = \phi_1(t_1)^{f(\beta_2,...,\beta_n)}
{\cal{S}}(\beta_2,...,\beta_n) 
\end{equation}
with ($i>1$) $\beta_i=\phi_i(t_i)/\phi_1(t_1)$ 
As before, $f(x_1,...,x_{n-1})$, and ${\cal{S}}(x_1,...,x_{n-1})$ 
are undetermined functions. 

\subsection{Some examples}

In many physical cases we might generally expect that 
the two times, $t$ and $t'$, 
scale in the same way, i.e., $\phi_1=\phi_2\equiv \phi$.
Below we explicitly list few interesting examples in this 
category.

A simple situation corresponds to 
a scaling function, $\phi(t)$, which is asymptotically 
a power law in $t$ (see ref.s in \cite{CM}),
i.e., one has $\phi(t)\sim t^{1/z}$, and the scaling of $C(t,t')$ is:
$C(t,t') = t^{f(t'/t)}{\cal{S}}(t'/t)$.
In most cases we expect the exponent $f$ to be a constant, so that:
\begin{equation}
C(t,t') = t^{f/z}{\cal{S}}(t'/t) ~ .
\label{sca_1}
\end{equation}
Asymptotically, the ``power scaling" of eq.~(\ref{sca_1}), 
is found in several 
toy models for glasses as in a ``phase space" model \cite{trap_m}, 
the Backgammon entropic barriers model \cite{Ritort}, 
the Queens long range interactions model \cite{Parisi_Queen},
in solvable models of interacting particles in high dimensionality 
\cite{CKP_sol}, or in a kinetically constrained lattice 
gas \cite{Sellitto} (see also references in \cite{BCKM}). 
Several of these cases are characterized by 
$f=0$. But in general one might expect cases with $f$
different from $0$. For instance in the Bak-Sneppen SOC model, 
the two time function $P(t,t')$, describing the return of activity
to a site at time $t$ which was most recently active at time $t'$, for an
avalanche started at $t=0$, seems to have a scaling of the form:
$P(t,t')=t^{z_{BS}}{\cal{P}}(t'/t)$ \cite{BoPa}, with constant $z_{BS}$. 
In a model of direct polymers in a random media a similar behavior is 
found for the off equilibrium ``overlap function": 
$q(t,t')=t^{-x}{\tilde{q}}(t'/t)$ \cite{Yoshino}.
In some models of non-linear diffusion 
equations \cite{stariolo}, correlation functions have been shown to have
a ``power law" scaling structure of eq.~(\ref{sca_1}) 
with constant exponents $f$ and with a scaling function ${\cal{S}}$ which is 
itself a power law. In the framework of out equilibrium dynamics, 
phenomena as coarsening or, more generally, 
phase ordering kinetics in ``standard" Ginzburg-Landau magnets 
usually show correlation functions which are asymptotically 
characterized (see \cite{Bray}) by the above scaling of eq.~(\ref{sca_1}), 
which is often called {\em ``simple or full or naive aging"}. 
In some discussions of glassy relaxation also 
a more complex, {\em ``interrupted aging"}, scenario was 
proposed \cite{BCKM,trap_m}, in which the long time regime of the 
two time autocorrelation function scales as 
$C(t,t')={\cal{S}}(t'/t^{1+\mu})$. In the present 
picture this corresponds to two different power exponents 
for $\phi_1(t)\sim t^{1/z}$ and $\phi_2(t')\sim t'^{1/z+\mu}$. 

In the case where $f=f(t'/t)$ is a non-trivial function of the 
scaling variable $\beta=t'/t$, one finds a {\em multiscaling} 
dynamical behavior. This is analogous to the multiscaling 
found, in different context, 
by Coniglio and Zannetti in the spinodal decomposition
of the $N=\infty$ Ginzburg Landau model with conserved order parameter 
or the one proposed also for the density profile of the DLA model \cite{CZ}. 

\medskip 

In the previous cases, the scaling variable was the ratio of powers of the 
two involved times, however in different situations, as for instance 
in the limit in which the exponent $1/z$ goes to zero, one may 
expect to have a logarithmic behavior for $\phi$: $\phi(t)\sim \ln(t)$. 
This situation gives as scaling structure:
$C(t,t') = \ln(t)^{f(\ln(t')/\ln(t))}{\cal{S}}(\ln(t')/\ln(t))$. 
In many cases one has $f=0$, namely:
\begin{equation}
C(t,t') = {\cal{S}}(\ln(t')/\ln(t)) ~ .
\label{sca_2}
\end{equation}
The {\em ``logarithmic scaling"} of eq.~(\ref{sca_2}) is found in several 
systems. An example of diffusion which shows the ``logarithmic scaling" 
is the one dimensional Sinai's model with a random local bias. 
In this case, for instance, the two time residency probability 
asymptotically has a scaling form given by  
eq.(\ref{sca_2}) with a scaling function ${\cal{S}}(\beta)$ 
which is an exponential corrected by a power law in $\beta=\ln(t')/\ln(t)$ 
\cite{FLDM}. 
Random Field systems also show ``logarithmic scaling" \cite{Bray,BCKM}, 
but also experimental random exchange Ising 
ferromagnets \cite{Schins}, among many other \cite{Bray,BCKM}, belong 
to this category.
Logarithmic kinetics have been also recently experimentally observed 
in the amorphous-amorphous transformations in some glasses under high 
pressure \cite{AAT}.
Interestingly, also non-thermal systems as granular media, shaken at low 
vibration amplitudes, present a non-trivial 
out of equilibrium dynamics, where numerical calculations on 
different models \cite{NC_hyst} 
suggest a ``logarithmic" scaling in the relaxation of the two time 
density correlation function as in eq.(\ref{sca_2}). 

The scenario about other disordered systems such as 
spin glass models is still controversial. 
To describe numerical calculations and to fit experimental data 
of relaxation in the thermoremanent magnetization of some spin glasses 
several proposals as power scaling eq.(\ref{sca_1}) or logarithmic scaling 
eq.(\ref{sca_2}) have been made \cite{FH,Marinari,Rossetti,Rieger,BCKM}. 
Also in recent computer simulations 
of a Lennard-Jones off equilibrium glass model 
the asymptotic behavior of the autocorrelation function was suggested to 
have a ``logarithmic" scaling \cite{Mussel-Rieger} as opposite to 
a ``power law" scaling previously proposed \cite{Barrat}. 

\section{The ``mixing" case}

Up to now we have dealt with {\em ``non mixing"} scale 
transformations, as in eq.~(\ref{gen_tra_t}), where the 
scaling of each of the variables doesn't depend on the other. 
However, situations where {\em ``mixing''} is present might be possible. 
Formally the case of mixing may be dealt with as the non-mixing one, 
however the results are too general to be of immediate practical use. 
For sake of completeness, we just show them. 
In the mixing case one finds that eq.~(\ref{sca_fin}) must be 
substituted by: $C(t,t') = \phi_1(t,t')^{f(\beta)}{\cal{S}}(\beta)$,
where the scaling variable, $\beta$, is now:
$\beta=\phi_2(t,t')/\phi_1(t,t')$.
Here, as before, the $\phi_i$ ($i=1,2$) are two unknown functions 
fixed by the mixing transformations 
$\tilde{t}=F_1(t,t',l)$ and $\tilde{t'}=F_2(t,t',l)$
(with $F_1(t,t',1)=t$ and $F_2(t,t',1)=t'$). 
The above result may be of scarce use because any function 
of two variables, $C(t,t')$, may be written as above 
in terms of other two functions, $\phi_1(t,t')$ and $\phi_2(t,t')$. 

However, it may be interesting to work out a specific example 
of mixing transformations, which shows how one may recover, from  
simple scale principles, a multifractal scaling structure. 

\subsection{An example of mixing}

While the function $C(t,t')$ has the general scaling property 
of eq.~(\ref{gen_sca_C}), we now suppose that the rescaling 
transformations of $t$ and $t'$ have the following specific form 
under a scale change of extension $l$:
\begin{equation}
\tilde{t}=t/l ~~~~~~  \tilde{t'}=t'/l^{z(t,t')} ~~ .
\label{spe_tra_t}
\end{equation}

Interestingly, within this context, we find that the scaling of $C(t,t')$ 
is restricted to have the following structure:
\begin{equation}
C(t,t') = t^{f(\beta)}{\cal{S}}(\beta) ~~ .
\label{sca_fin_spe}
\end{equation}
Here the scaling variable, $\beta$, has only two possible forms:
either it is a ratio of powers of the two times 
\begin{equation}
\beta=t'/t^z
\label{power_sca_spe}
\end{equation}
with $z=const$ (corresponding to a non-mixing case previously described), or 
\begin{equation}
\beta=\frac{\ln(t')}{\ln(t)}+\frac{H(\beta)}{\ln(t)}
\label{log_sca}
\end{equation}
where $H(\beta)$ is an undetermined function. 

The scaling form  (\ref{power_sca_spe}) corresponds to the case  
$z(t,t')=const$, which is one of the non-mixing cases we dealt with before. 
The scaling form given in eq. (\ref{log_sca}) corresponds 
instead to a non constant scaling exponent 
$z=z(t,t')$ in (\ref{gen_tra_t}), which thus gives a mixing transformation
of $t$ and $t'$.
Actually, it turns out that the only possible solution for a non constant $z$
is $z(t,t')=\beta$ with $\beta$ given in eq.~(\ref{log_sca}).
In this case the scaling variable is asymptotically 
logarithmic in the two times, $\beta=\ln(t')/\ln(t)+O(1/\ln(t))$. 
This kind of scaling for different variables was proposed, for instance, 
for the multifractal description in turbulence \cite{mandel,turbulence}, 
in the DLA model\cite{amitrano}, in Self-Organized-Critical (SOC) 
models \cite{Kadanoff_ms} or in voltage 
distribution of random resistor networks \cite{CM,deArcangelis}.
These scaling forms, differently from ordinary critical phenomena,  
are characterized by a continuity of scaling exponents. 

For definiteness it is interesting to work out the simple case 
where the function $H(\beta)$ is linear in $\beta$, 
situation which might generically correspond to the case of very long 
times $t$ and small $\beta$. By writing  
$H(\beta)=\beta\ln(t_0)-\ln(t'_0)$ (where $t_0$ and $t'_0$ are constants), 
from eq.~(\ref{log_sca}) 
one finds that $\beta=\ln(t'/t'_0)/\ln(t/t_0)$. This case 
corresponds for instance to the multifractal scaling proposed by 
Kadanoff et al. in Ref.\cite{Kadanoff_ms} to describe the avalanches size
distribution in the context of SOC models. 

Below we work out the example of mixing transformation of 
eq.~(\ref{spe_tra_t}) in details. 
As in the general case above, we have to impose the group rules 
on the scale transformations.
For the transformation of the variable $t$ and $t'$ this implies
($i=1,2$):
\begin{equation}
F_i(F_1(t,t',l_1),F_{2}(t,t',l_1),l_2)=F_i(t,t',l_1\cdot l_2) ~~ .
\end{equation}
For a transformation as in eq.~(\ref{spe_tra_t}), this assertion simply 
imposes that:
\begin{equation}
\frac{d}{d l} z(t(l),t'(l)) = 0
\label{eq_z}
\end{equation}
The above eq.~(\ref{eq_z}) has two kind of solution. The first is the 
trivial one: $z=const.$.
The second is non-trivial and has the following form:
\begin{equation}
z=\frac{\ln(t')}{\ln(t)}+\frac{H(z)}{\ln(t)} ~~ , 
\label{sol_log}
\end{equation}
where $H(z)$ is a generic function. 
The latter may be obtained as follows.
Since the function $z(\tilde{t},\tilde{t'})\equiv z(t/l,t'/l^z)$ is invariant 
under rescaling eq.~(\ref{eq_z}), we can write that $z(t/l,t'/l^z)=z(t,t')$. 
By fixing $l=t$, we obtain : $z(t,t')=z(1,t'/t^z)$. 
Thus $z$ is a function of the single variable $t'/t^z$, and we can write 
$z(t,t')=g(t'/t^z)$. Here we have defined 
\begin{equation}
g(t'/t^z) \equiv z(1,t'/t^z) ~~ .
\end{equation}
By inverting the above 
relation we have: $t'/t^z=g^{-1}(z)$, and passing to the logarithms we 
recover $z=\ln(t')/\ln(t)+H(z)/\ln(t)$, where we have introduced 
the unknown generic function $H(z)=-\ln(g^{-1}(z))$. Thus we have found the 
solution given in eq.~(\ref{sol_log}). This result states that, for fixed $z$, 
whenever $\ln(t')$ and $\ln(t)$ are large enough, we have 
$z=\ln(t')/\ln(t)+O(1/\ln(t))$. 

By imposing then group properties to the function $C(t,t')$ itself
(see eq.~(\ref{group_on_C})), we obtain eq.~(\ref{eq_f}), which, after 
insertion of eq.~(\ref{spe_tra_t}), implies that $f(t/l,t'/l^z)=f(t,t')$, 
and by taking $l=t$, as above, we have that 
$f(t,t')=f(1,t'/t^{z(t,t')})$. 

Whenever $z$ is a constant 
we recover a non-mixing case described in the previous section.
Let's now suppose that $z$ is not a constant and is given 
by eq.~(\ref{sol_log}). In this case, 
we can prove that the exponent $f(t,t')$ is 
a function of the single variable $z$: $f(t,t')=f(z)$. In fact, 
as before we have that $f(t,t')=f(1,t'/t^z)=f(1,g^{-1}(z))$, i.e., 
$f$ is a function of the variable $z$. Analogously one proves that 
$C(t,t')= t^{f(z)} {\cal{S}}(z)$, where 
the scaling function ${\cal{S}}$ is now ${\cal{S}}(x)=C(1,g^{-1}(x))$. 
From this result, in the asymptotic limit of large $t'$ and $t$, 
we recover the scaling form for $C(t,t')$, given in 
eq.s~(\ref{sca_fin_spe}) and (\ref{log_sca}). 

\section{Conclusions}

We expect that the present approach may be useful to describe general 
properties of dynamical functions in physical systems when their 
characteristic times diverge, since in such a situation, like close to usual 
critical points, scale invariance should be reasonably present. 
Actually, one observes diverging characteristic times 
typically when out of equilibrium dynamics phenomena become 
important, i.e., when  an explicit dependence of functions as 
$C(t,t')$ on both times (and not on their difference) is observed, 
a fact which is some time generically called ``aging". In this perspective 
the structural properties of scaling described here should be 
naturally associated to out of equilibrium dynamics 
(i.e., to ``aging") effects.
Interestingly, we have pointed out that 
in a broad variety of physical systems, ranging from magnets, to polymers, 
glasses, or spin glasses, random fields, random ferromagnets, granular materials, 
diffusive systems, etc..., 
one observes scaling properties of dynamical functions which may be 
well inserted in the framework reported above. 

We have shown in full generality that a generalized homogeneous function, 
$C(t,t')$, which acts as in eq.~(\ref{gen_sca_C}) under the scale 
transformation of its variables given in eq.~(\ref{gen_tra_t}), must obey 
the scaling behavior of eq.~(\ref{sca_fin}). In this theoretical 
framework, also a multiscaling or multifractal behavior 
is admissible in the dynamics. It would 
be interesting to understand if it exists in real dynamical systems. 

The present approach is not restricted to scaling of 
dynamical functions. We have seen that it is relevant to 
describe, as well known, usual scaling in standard critical phenomena,  
but it also describes multiscaling or 
multifractal properties introduced in apparently 
completely different systems as, for instance, models of 
Self-Organized-Criticality, DLA, random resistor networks.
In this sense this approach 
may help to rationalize the existence of very few broad 
``universality classes" found in the scaling behaviors in 
very different contexts. 

\bigskip

This work was partially supported by the TMR Network Contract ERBFMRXCT980183 
and from MURST (PRIN-97).


\begin{thebibliography}{40}

\bibitem{HH} P.C. Hohenberg and B.I. Halperin, Rev. Mod. Phys. {\bf 49},
435 (1977).

\bibitem{Bray} A. J. Bray, Adv. Phys. {\bf 43}, 357 (1994).

\bibitem{BCKM} J.P. Bouchaud, L.F. Cugliandolo, J. Kurchan and M. Mezard,
in {\em Spin glasses and random fields}, A.P. Young ed.
(World Scientific, 1997).
E. Vincent, J. Hammann, M. Ocio, J.-P. Bouchaud, L.F. Cugliandolo,
1997 {\em Sitges Conf. on Glassy Systems} ed. M. Rubi (Berlin:
Springer).

\bibitem{CM} A. Coniglio and M. Marinaro, Physica {\bf 54}, 261 (1971).
A. Coniglio, Physica A {\bf 140}, 51 (1986).

\bibitem{mandel}
B.B. Mandelbrot, J. Fluid. Mech. {\bf 62}, 331 (1974). 

\bibitem{turbulence}
G. Parisi and U. Frish, in the proceed. of the {\em International School 
of Physics ``E. Fermi"}, Course LXXXVIII, eds. M. Ghil, R. Benzi, G. Parisi 
(North-Holland, Amsterdam, 1985).
U. Frish, M. Vergassola, Europhys. Lett. {\bf 14}, 439 (1991). 

\bibitem{amitrano} C. Amitrano, A. Coniglio, F. di Liberto, 
Phys. Rev. Lett. {\bf 57}, 1016 (1986). 

\bibitem{Kadanoff_ms} L.P. Kadanoff, S. Nagel, L. Wu, S. Zhou, 
Phys. Rev. A {\bf 39}, 6524 (1989). 

\bibitem{deArcangelis} L. de Arcangelis, S. Redner and A. Coniglio, 
Phys. Rev. B {\bf 31}, 4725 (1985). 

\bibitem{halsey} T.C. Halsey, M.H. Jensen, L.P. Kadanoff, I. Procaccia,
B.I. Shraiman, Phys. Rev. A {\bf 33}, 1141 (1986).

\bibitem{PV} see for example the review article 
G. Paladin and A. Vulpiani, Phys. Rep. {\bf 156}, 147 (1987). 

\bibitem{Stanley} H.E. Stanley and P. Meakin, Nature {\bf 335}, 405 (1988). 

\bibitem{CZ} A. Coniglio and M. Zannetti, Physica D {\bf 38}, 37 (1989).

\bibitem{nota1}
Here we consider $t'$ and $t$ as the proper scaling variables, 
but more generally one might take linear combinations (as $t-t'$ and $t'$).

\bibitem{Cu_Ku} L.F. Cugliandolo and J. Kurchan, J. Phys. A {\bf 27}, 
5749 (1994).

\bibitem{trap_m} J.P. Bouchaud, J. Physique (France) {\bf 2}, 1705 (1992); 
J.P. Bouchaud and D.S. Dean, J. Physique I (France) {\bf 5}, 
265 (1995).

\bibitem{NC_hyst} M. Nicodemi and A. Coniglio, 
Phys. Rev. Lett. {\bf 82}, 916 (1999). M. Nicodemi, cond-mat/9809346.

\bibitem{Ritort} F. Ritort, Phys. Rev. Lett. {\bf 75}, 1190 (1995).

\bibitem{Parisi_Queen} D.S. Dean and G. Parisi, preprint cond-mat/9711057.

\bibitem{CKP_sol} L.F. Cugliandolo, J. Kurchan, G. Parisi, 
Phys. Rev. Lett. {\bf 74}, 1012 (1995).

\bibitem{Sellitto} J. Kurchan, L. Peliti, M. Sellitto, Europhys. Lett. 
{\bf 39}, 365 (1997). L. Peliti and M. Sellitto, preprint cond-mat/9712221.

\bibitem{BoPa} S. Boettcher and M. Paczuski, Phys. Rev. Lett. {\bf 79}, 
889 (1997).

\bibitem{Yoshino} H. Yoshino, Jour. Phys. A {\bf 29}, 1421 (1996). 

\bibitem{stariolo} D.A. Stariolo, Phys. Rev. Lett., in press; 
cond-mat/9612082. 

\bibitem{FLDM} D.S. Fisher, P. Le Doussal and C. Monthus, Phys. Rev. Lett. 
{\bf 80}, 3539 (1998). 

\bibitem{Schins} A.G. Schins, A.F.M. Arts, H.W. de Wijn, 
Phys. Rev. Lett. {\bf 70}, 2340 (1993).

\bibitem{AAT} O.B. Tsiok, V.V. Brazhkin, A.G. Lyapin, L.G. Khvostantev,
Phys. Rev. Lett. {\bf 80}, 999 (1998).

\bibitem{FH} D.S. Fisher and D.A. Huse, Phys. Rev. Lett. {\bf 56}, 1601
 (1986); Phys. Rev. B {\bf 38}, 373 and 386 (1988). 

\bibitem{Marinari} E. Marinari, G. Parisi, J.J. Ruiz-Lorenzo, 
in {\em Spin glasses and random fields}, A.P. Young ed.
(World Scientific, 1997); cond-mat/9701016.

\bibitem{Rossetti} E. Marinari, G. Parisi, D. Rossetti, preprint 
cond-mat/9708025.

\bibitem{Rieger} H. Rieger, {\em Annual Rev. Comp. Phys.}, 
vol.II ed. D. Stauffer (Singapore: World Scientific), p.295. 

\bibitem{Mussel-Rieger} O. Mussel and H. Rieger, preprint cond-mat/9804063.

\bibitem{Barrat} J.L. Barrat and W. Kob, Phys. Rev. Lett. {\bf 78}, 
4581 (1997); preprint cond-mat/9804103.

\end{thebibliography}
\end{document}